\documentclass[trackchanges]{aastex631}

\shorttitle{Unveiling the Fermi Bubbles origin with MeV photon telescopes}
\shortauthors{M. Negro et al.}
\graphicspath{{./}{figures/}}
\begin{document}

\title{Unveiling the Fermi Bubbles origin with MeV photon telescopes}

\author[0000-0002-6548-5622]{Michela Negro}
\affiliation{University of Maryland, Baltimore County, Baltimore, MD 21250, USA}
\affiliation{NASA Goddard Space Flight Center, Greenbelt, MD 20771, USA}
\affiliation{Center for Research and Exploration in Space Science and Technology, NASA/GSFC, Greenbelt, MD 20771, USA}

\author[0000-0002-0794-8780]{Henrike Fleischhack}
\affiliation{Catholic University of America, 620 Michigan Ave NE, Washington, DC 20064, USA}
\affiliation{NASA Goddard Space Flight Center, Greenbelt, MD 20771, USA}
\affiliation{Center for Research and Exploration in Space Science and Technology, NASA/GSFC, Greenbelt, MD 20771, USA}

\author[0000-0001-9067-3150]{Andreas Zoglauer}
\affiliation{Space Sciences Laboratory, University of California at Berkeley, 7 Gauss Way, Berkeley, CA 94720, USA}

% \collaboration{6}{(AAS Journals Data Editors)}
\author[0000-0002-5296-4720]{Seth Digel}
\affiliation{KIPAC/SLAC, 2575 Sand Hill Road, Menlo Park, CA  94025}

\author[0000-0002-6584-1703]{Marco Ajello}
\affiliation{Department of Physics and Astronomy, Clemson University, Clemson, SC 29634, USA}

%% Note that the \and command from previous versions of AASTeX is now
%% depreciated in this version as it is no longer necessary. AASTeX 
%% automatically takes care of all commas and "and"s between authors names.

%% AASTeX 6.31 has the new \collaboration and \nocollaboration commands to
%% provide the collaboration status of a group of authors. These commands 
%% can be used either before or after the list of corresponding authors. The
%% argument for \collaboration is the collaboration identifier. Authors are
%% encouraged to surround collaboration identifiers with ()s. The 
%% \nocollaboration command takes no argument and exists to indicate that
%% the nearby authors are not part of surrounding collaborations.

%% Mark off the abstract in the ``abstract'' environment. 
\begin{abstract}

The Fermi Bubbles (FB) are a pair of large-scale ellipsoidal structures extending above and below the Galactic plane almost symmetrically aligned with the Galactic Center. After more than 10 years since their discovery, their nature and origin remain unclear. Unveiling the primary emission mechanisms, whether hadronic or leptonic, is considered the main tool to shed light on the topic. We explore the potential key role of MeV observations of the FB and we provide a recipe to determine the sensitivity of Compton and Compton-pair telescopes to the extended emission of the FB. %define the main requirements for future gamma-ray telescopes operating in this regime in order to be able to disentangle the different emission mechanism scenarios. We 
We illustrate the capabilities of the Imaging Compton Telescope COMPTEL, the newly selected NASA MeV mission COSI (Compton Spectrometer and Imager), as well as the expectations for a potential future Compton-pair telescope such as AMEGO-X (All-sky Medium Energy Gamma-ray Observatory eXplorer).
\end{abstract}

%% Keywords should appear after the \end{abstract} command. 
%% The AAS Journals now uses Unified Astronomy Thesaurus concepts:
%% https://astrothesaurus.org
%% You will be asked to selected these concepts during the submission process
%% but this old "keyword" functionality is maintained in case authors want
%% to include these concepts in their preprints.

%% From the front matter, we move on to the body of the paper.
%% Sections are demarcated by \section and \subsection, respectively.
%% Observe the use of the LaTeX \label
%% command after the \subsection to give a symbolic KEY to the
%% subsection for cross-referencing in a \ref command.
%% You can use LaTeX's \ref and \label commands to keep track of
%% cross-references to sections, equations, tables, and figures.
%% That way, if you change the order of any elements, LaTeX will
%% automatically renumber them.
%%
%% We recommend that authors also use the natbib \citep
%% and \citet commands to identify citations.  The citations are
%% tied to the reference list via symbolic KEYs. The KEY corresponds
%% to the KEY in the \bibitem in the reference list below. 

\section{Introduction} \label{sec:intro}

More than ten years have passed since the \textit{Fermi} bubbles (FB) were discovered \citep{SuFBdiscovery, FBdiscovery}. These Galactic-scale structures extending, almost symmetrically, above and below the Galactic plane, have been observed, analyzed, and physically interpreted in numerous and diverse studies, and yet, their nature, their origin and their emission mechanisms are still under investigation.
Detections of the emission of the FB peaks in the GeV $\gamma$-ray band, domain of the \textit{Fermi} Large Area Telescope (LAT) since 2008 \citep{theLAT}. \cite{FBLAT} have provided the detailed measurement and analysis of the FB between 100 MeV and 500 GeV beyond 10 degrees from the Galactic plane. Analyses have revealed that the intensity spectrum is almost spatially uniform within the sharp edges of the bubbles \citep{FBdiscovery, Dobler2012, Ade2013, FBLAT}, and it is well represented by a power law with a hard photon spectral index ($\sim-2$) and an exponential cutoff at high energies ($\sim$110 GeV). The FB interestingly overlaps (at $|b|<35^\circ$) with another long known \textit{elephant}\footnote{Elephants in the gamma-ray sky: \url{ https://cerncourier.com/a/elephants-in-the-gamma-ray-sky/}} in the sky: the so-called microwave haze, observed by WMAP\footnote{\url{https://map.gsfc.nasa.gov}} \citep{Finkbeiner, Rubtsov:2018eqv}. The similarity of their shapes is impressive, making the association between the two giants very straightforward. 

At latitudes within 10 degrees from the Galactic plane, \cite{MalyshevHerold} found  GeV intensities than greater than at high latitudes with a spectrum well described by a single power law between 10 GeV and 1 TeV and, interestingly, the axis perpendicular to the plane of the Milky Way was found to be shifted to the west of the Galactic Center (GC), which undermines the generally accepted hypothesis that the origin of the FB is associated with the GC. 

The observation of the FB at TeV energies is the territory of ground-based $\gamma$-ray observatories. So far, no significant emission has been observed in that energy range.  The High Altitude Water Cherenkov (HAWC) observatory\footnote{\url{https://www.hawc-observatory.org/}} derived stringent upper limits on the VHE gamma-ray emission from the the northern FB (outside the Galactic plane) above 1.2 TeV \citep{HAWCuplim}. These measurements provide further support for an exponential cutoff in the energy spectrum and rule out the presence of a second component at very high energies. 

There have been some indications that the spectrum of the FB nearby the GC may be harder, even present a feature peaking between $1-4$ GeV \citep{HooperSlatyer} and could extend to higher energies compared to the rest of the bubbles \citep{MalyshevHerold, latGC}. However, such results suffer from large systematic uncertainties due to the complicated modeling of the interstellar diffuse and discrete source gamma-ray emission of GC region, which may result in contamination from non-modelled components (e.g., unresolved pulsar wind nebulae \citep{vecchiotti2021contribution}). A recent study by the H.E.S.S. Collaboration \footnote{\url{https://www.mpi-hd.mpg.de/hfm/HESS/}} found no significant emission from the low-latitude part of the bubbles, constraining the exponential cutoff in the energy spectrum to be between 700 GeV to 1.7 TeV \citep{HESS_FB}.

The eROSITA telescope \citep{eROSITA} recently reported an X-ray counterpart, the \textit{eROSITA bubbles} (eRB) \citep{eRBNature}. The eRB have  morphology different from the FB: they are larger, more spherical, and they show brighter edges with respect of the inner regions. It has been argued that eRB and FB must be related due to their positional concentricity. The observed average X-ray surface brightness reported in \cite{eRBNature} is almost two orders of magnitude above the FB gamma-ray intensity\footnote{The eRB average X-ray surface brightness between 0.6 and 1 keV, as quoted in \cite{eRBNature}, is $(2-4) \times$10$^{-15}$ erg cm$^{-2}$s$^{-1}$arcmin$^{-2}$, the two numbers in the range corresponding to the northern and the southern FB respectively.}, which roughly corresponds to  $(1.5-3) \times$10$^{4}$ eV cm$^{-2}$s$^{-1}$sr$^{-1}$. This mismatch might be indicative of different energy sources behind the two pairs of bubbles. More studies are needed to understand the connection, if any, between the FB and the eRB.

Where and how did the FB originate? Although measuring the distance of the FB along the line of sight is not straightforward, it is appealing to assume their coincidence with the GC, given their shape projected on the sphere. In this scenario, previous publications distinguish between an origin due to nuclear star-formation (NSF) activity near the GC, and past jet-like activity associated with the super-massive black hole in the center of the Milky Way. In both these scenarios, cosmic rays (CR) are produced and accelerated close to the Galactic plane and then transported further out by strong winds \citep{hYKYang2017}. Once the FB are inflated, two possible emission mechanisms can contribute to the observed GeV emission: hadronic and leptonic processes, depending on the nature of the main species of particles involved.

Hadronic processes involve cosmic rays protons (CRp) interacting with thermal nuclei in the vicinity of the GC, producing short-lived mesons such as neutral and charged pions. The majority of the observed GeV $\gamma$-ray emission would be associated with $\pi^0$ decay. However, other decay products of hadronic mechanisms comprise a population of secondary leptons, including electrons and positrons (CRe) which produce additional gamma rays via inverse Compton (IC) scattering, and high-energy neutrinos. The latter are indeed expected to contribute to the astrophysical neutrino flux detected by the high-energy neutrino observatories, such as IceCube\footnote{\url{https://icecube.wisc.edu}} and ANTARES\footnote{\url{https://antares.in2p3.fr}}. Recent works have derived upper limits on the neutrino flux from the FB considering both the LAT observation and the HAWC upper limits \citep{keFengFBnu, Razzaque2018, ANTARESnuFB}. One advantage of hadronic transport models is that the diffusion from the GC to the bubbles' edges can proceed with no significant energy losses, which is consistent with a long life for the FB of the order of $\sim$Gyrs \citep{CrockerHadronic, FBLAT, keFengFBnu}. Pure hadronic scenarios, however, fail to reproduce the microwave haze observations: the synchrotron component produced by the secondary e$^+$e$^-$ population would be too soft and 3--4 times lower than the observed emission. In order to reproduce WMAP data, it is necessary to invoke a primary source of CRe or re-acceleration mechanisms for the secondary population.

Leptonic processes involve CRe interacting with the low-energy photons of the interstellar radiation field (ISRF). In this scenario, the GeV $\gamma$-ray emission could be entirely attributed to IC scattering, and the lower-energy part of the CRe population can also naturally explain (fine tuning the assumed average magnetic field within the FB) the microwave haze emission via synchrotron emission. The Bremsstrahlung component from the CRe is negligible due to the low gas density in the region of the FB (in literature it is often quoted as n$_H=0.01$ cm$^{-3}$). Apart from reproducing the haze observations, leptonic scenarios also have the advantage that they easily explain the high-energy cutoff observed by the LAT: very high energy CRe (above tens of TeV) cool much more efficiently via IC and synchrotron energy losses \citep{hYKYang2017}. However, pure leptonic models set a stringent limit on the age of the FB, because a CRe of TeV energies radiating due to both the Galactic magnetic field (via synchrotron) and the ISRF (via IC) exhausts its energy in a few hundred-thousand years. This limits the age of the FB to be relatively young. Particles would need to be promptly accelerated and transported to the edges of the FB by extremely fast winds (like the jets observed in active galactic nuclei), unless  re-acceleration processes at recent times are invoked \citep[see, e.g.,][]{KSCheng2014, PMertsch2011}. It has been pointed out that the fast cooling of CRe is difficult to rectify with the observed uniformity of the GeV emission throughout the FB: given the ISRF gradient  with Galactic latitude, the energy density of the CRe would need to compensate for that in order to match the observed uniform brightness \citep{hYKYang2017}.

Hybrid models, where populations of CRe and CRp coexist within the FB, can also be used to explain the observations throughout the electromagnetic spectrum of the FB \citep{keFengFBnu}. A two-zone hybrid model, with CRe populating the lower part of the bubbles and the CRp inhabiting the cocoons, is also appealing and worth considering for future investigations.

In summary: As of now, both leptonic and hadronic scenarios can explain the observed GeV emission observed by the LAT, with different strengths and weaknesses for the two cases. HAWC's high-energy $\gamma$-ray upper limits are not stringent enough to rule out a hadronic component and high-energy neutrino observatories are not sensitive enough (and will not be soon: \cite{keFengFBnu}) to unambiguously detect neutrinos from the FB, which would  confirm a hadronic emission scenario. 

\begin{figure}[t]
    \centering
    \includegraphics[scale=0.45]{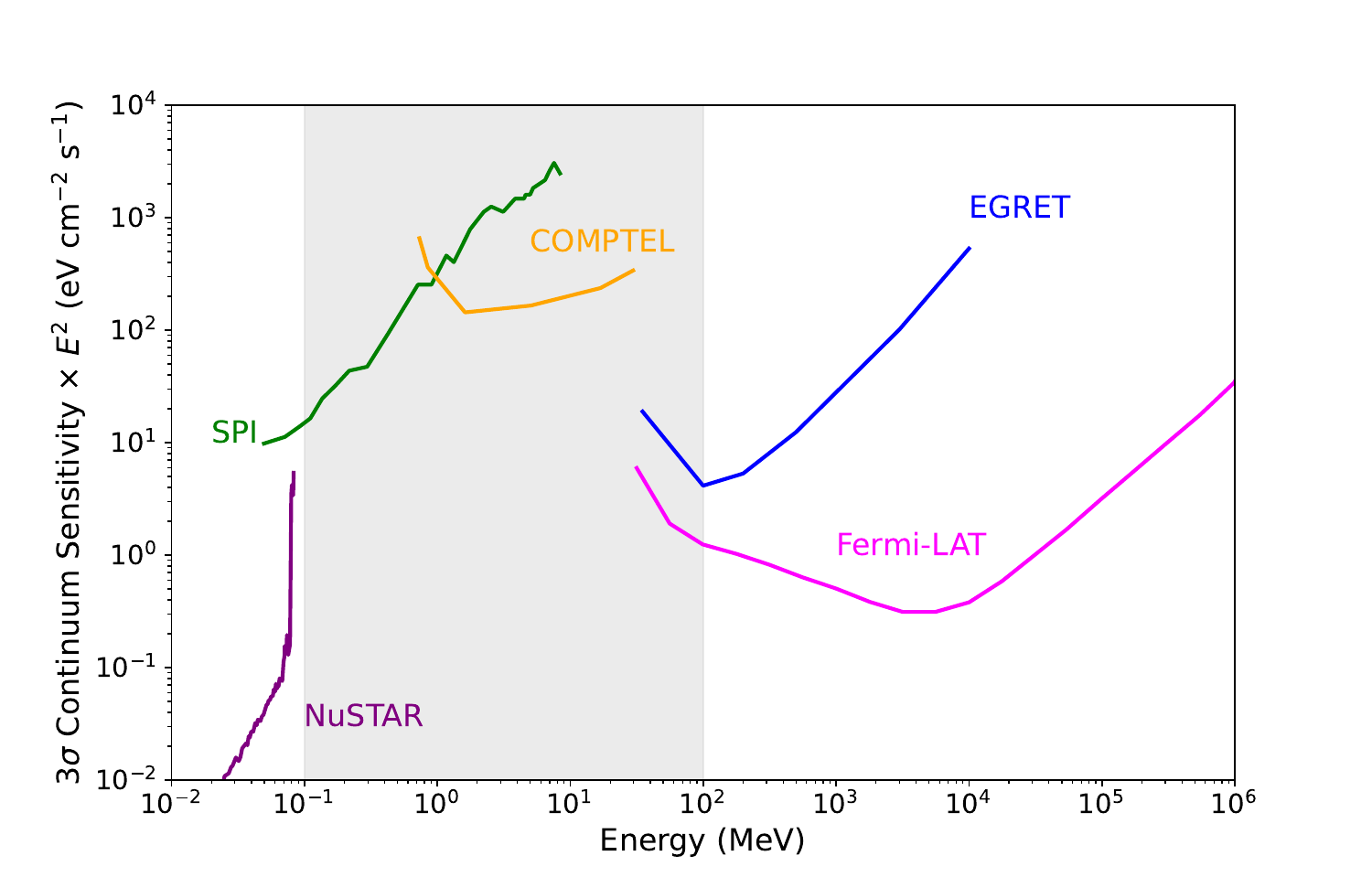}
    \caption{Sensitivities of past and present instruments operating around and within the MeV gap (gray shaded region). }
    \label{fig:mevgap}
\end{figure}

\begin{figure}[t]
    \centering
    \includegraphics[scale=0.45]{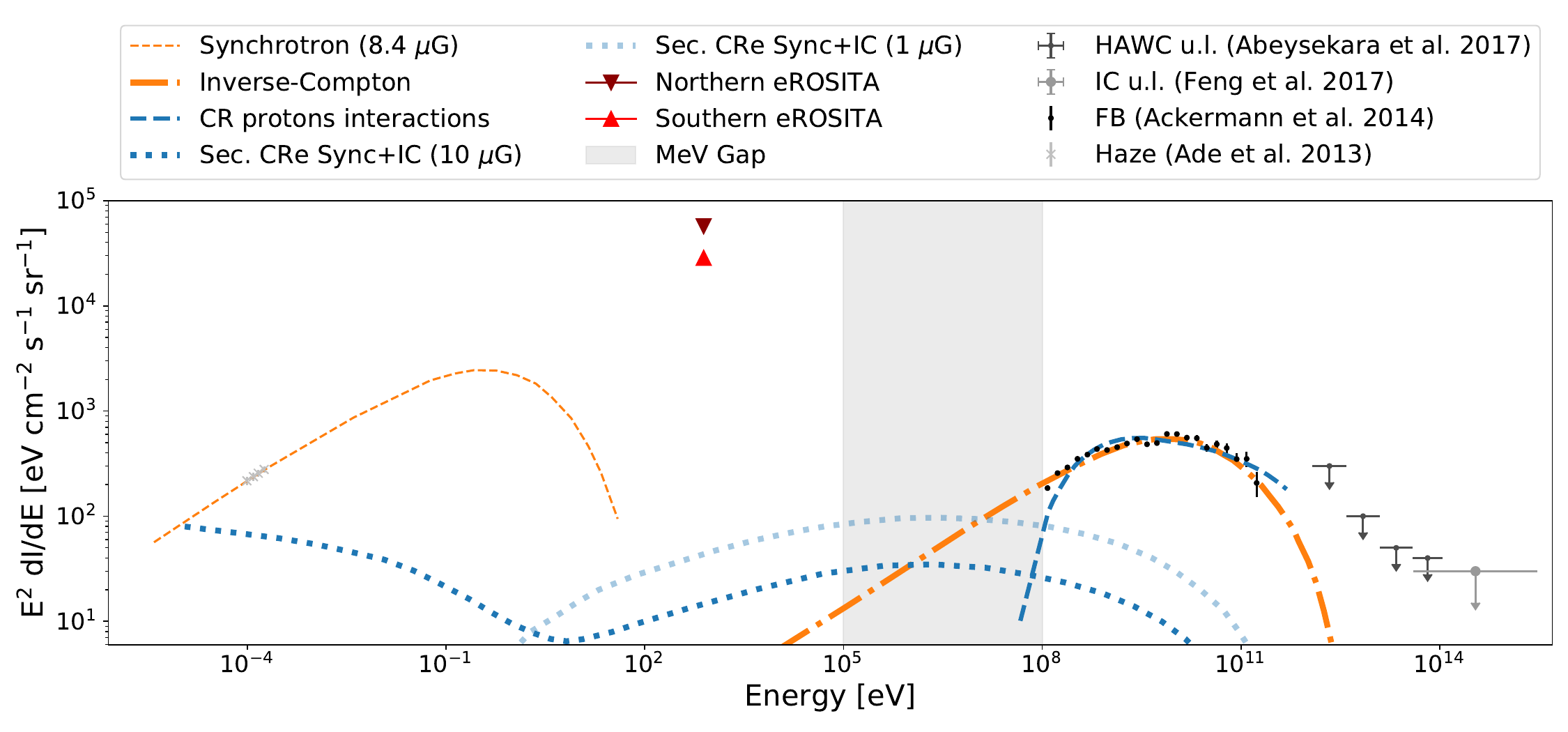}
    \caption{Compilation of spectral measurements of the FB together with models for the emission components.  The orange lines show the IC and synchrotron emission from the same benchmark population of electrons \citep{FBLAT} obtained from the best-fit to FB Fermi-LAT measurement (black points) and Planck microwave haze measurement \citep{Ade2013} (silver crosses). The blue lines show primary and secondary $\gamma$-ray emission from hadronic processes. Also shown are the HAWC and IceCube upper limits for very-high-energy gamma-rays and neutrinos respectively, and the average surface brightness of the eRB (red points). The shaded gray region marks the MeV gap.}
    \label{fig:landscape}
\end{figure}

If we look at currently operating high-energy telescopes from X-rays to very-high-energy gamma rays, we would notice a `gap' in sensitivity between $\sim 100$ keV and $\sim100$ MeV, with the worst performance around $1-10$ MeV. We illustrate the gap in sensitivity in the MeV band in Fig.~\ref{fig:mevgap}. This poorly explored band of the electromagnetic spectrum is called the \textit{MeV gap}. The number and variety of possibilities that observations in the \textit{MeV gap} can open for the astrophysical community are simply extraordinary \citep[see, e.g.][]{Andritschke2006, mcenery2019allsky, DeAngelis}, from the study of nuclear lines from supernovae, to the diffuse 511 keV emission possibly related to dark matter, to a whole class of Galactic and extragalactic astrophysical sources shining bright in this energy range. In particular, in this paper, we discuss the advantages that MeV observations of the FB would provide for unveiling the underlying emission mechanisms. In Fig.~\ref{fig:landscape} we highlight how within the MeV energy band the gamma-ray emission expected from the leptonic and the hadronic scenarios show remarkably different behaviours, which could be captured by a MeV telescope sensitive enough to detect the large-scale emission from the FB.

The paper is structured as follows: In Sec.~\ref{sec:MeVtelescopes} we discuss the detection and imaging principles for MeV $\gamma$-ray observations, and we provide some example of past, future and proposed MeV mission designs. In Sec.~\ref{sec:sensitivity} we present a procedure to compute the sensitivity of MeV telescopes to MeV emission from the FB, and illustrate the case for past, future and potential future Compton and Compton-pair telescopes. The last section, Sec.~\ref{sec:concl}, is devoted to the summary and the conclusions, and we include two appendices with details about the hadronic and leptonic models used in this work, and the scaling of the sensitivity in the Compton regime with the extension of a source. \\

\section{Measuring MeV $\gamma$-rays}
\label{sec:MeVtelescopes}
For the most part, the universe is transparent to hard X-rays and gamma rays below tens of GeV, which ironically also makes photons in this energy range challenging to detect. High-energy photons interact with matter in three different ways: below some tens to hundreds of keV (depending on the material), the dominant interaction mechanism is the photo-electric effect; at intermediate energies (tens of keV to tens of MeV), Compton scattering on quasi-free electrons has the highest cross section; above some tens of MeV, $e^+e^-$ pair production in the electric field of the ambient nuclei dominates.

\subsection{MeV gamma-ray detection principles}
In the MeV band the photo-electric effect is subdominant, and hence we focus here on Compton and pair-production regimes.

In the pair-production regime, the secondary $e^+e^-$ are typically energetic enough to propagate over macroscopic distances in the detector, leaving ionization tracks. For sufficiently high primary photon energies, the secondary particles may emit additional gamma rays interacting with the electric fields of ambient nuclei (Bremsstrahlung radiation), initiating a cascade or shower of electrons, positrons, and gamma rays. In the Compton regime, the scattered photon is emitted at a characteristic angle $\phi$ with respect to the primary photon direction and may undergo further Compton scattering processes before depositing the rest of its energy via the photoelectric effect. Depending on the energy transferred to the secondary electron, this particle may also travel over macroscopic distances and leave an ionization trail. In the following sections, Compton events for which the direction of the scattered electron is measured are referred to as ``tracked Compton events'' (TC events), and those without an electron track are called ``untracked Compton events" (UC events).
% \textcolor{blue}{Also the noise from the electronics can contribute to the instrumental background. ???}

While the interaction cross sections depend on the detector material, generally the cross sections for Compton scattering and pair production are several orders of magnitude lower than for the photoelectric effect. MeV gamma-ray detectors thus require a large enough mass and volume to have a reasonable chance that gamma rays will interact in the detector and not just pass through.

\subsection{Imaging MeV gamma rays}
For instruments designed to observe gamma-ray bursts, imaging capabilities are often unnecessary. For the duration of the burst, the gamma-ray signal outshines the background, and it is possible to provide a crude localization of the burst by comparing, e.g., the arrival times or count rates in different parts of the detector. However, imaging capabilities can be crucial for the performance of missions designed to study gamma-ray emission on longer time scales, allowing them to suppress the background, disentangle emission from multiple sources in the field of view, and study the morphology/substructure and extent of some gamma-ray sources.  

There is no known mechanism to focus MeV gamma rays with a technology sufficiently tested for space missions\footnote{Multi-layer Laue lenses \citep{MurrayLaueLenses}. Tungsten/silicon carbide multilayers with reduced interface roughness can focus X-rays above 100 keV and up to 250 keV \citep{WSiCxrayoptics, WCSiCxrayoptics}. However, this technology has not been tested in space-based applications.}.  ``Imaging'' is achieved by reconstructing the arrival directions of each incoming gamma-ray photon (on an event-by-event basis). This can only be achieved for Compton scattering and pair production events. 

For Compton events, it is typically possible to measure the direction of the scattered photon (from the location of the first and second interaction in the detector) and the Compton angle $\phi$ (from the deposited energies). The arrival direction of the primary photon projected back on the sky is then constrained to lie on a circle with opening angle $2\phi$ around the direction of the scattered photon. For imaging instruments, the angular resolution (uncertainty in the width of the Compton ring) is typically dominated by the uncertainty in $\phi$, which is determined by the energy resolution. The FWHM of the distribution of the difference between true and reconstructed Compton angle is often referred to as ARM (angular resolution measure).%\cite{refARM}.

Unambiguously measuring the arrival direction of the primary photon is not possible given just the scattered photon and Compton angle. Statistical methods exist to extract the number and likely location(s) of gamma-ray sources in the field of view, even in the presence of diffuse background, as long as a sufficient number of photons are observed from each source. 

TC events have a second constraint on the arrival direction of the primary photon, given by the ``scatter plane'': The primary gamma-ray direction must be co-planar with the directions of the Compton electron and the scattered photon. The uncertainty in the scatter plane is called ``scatter plane deviation'' (SPD). Due to uncertainty in the SPD, it generally reduces the Compton circle to an arc segment rather than a point. The scatter plane constraint helps reduce ambiguities and distinguish weak sources from chance overlap of Compton circles. 

%\textcolor{blue}{Typical ARM/SPD numbers? Andreas can help here.}
The angular resolution of a Compton telescope can be improved to as fine as a fraction of a degree by means of coded aperture masks \citep{CodedMasks, ComptonCodedMask}. 
However, the introduction of a coded aperture mask may also result in a reduction of the instrumental field of view, of the efficiency (fewer photons interacting in the detector), and potentially in an increased instrumental background due to activation of the mask materials. Furthermore, coded masks suppress diffuse emission on spatial scales larger than the typical solid angle covered by the mask pixels: this results in a improved sensitivity to point sources, but it is detrimental for detecting large-scale diffuse emission.

In the case of gamma rays interacting via $e^+e^-$ pair production (which we will address as ``P events''), the direction of the primary photon can be estimated from a combination of the tracks of the electron and the positron. The angular resolution is described by the point-spread function (PSF), the distribution of reconstructed arrival directions for gamma rays from a point source at large distance. The PSF is typically radially symmetric around the true direction, with a Gaussian peak but larger tails (e.g., the PSF of the \textit{Fermi} Large Area Telescope (LAT) \citep{theLAT} is well described by a King function \citep{Kingfunc}). The width of the distribution depends on the instrument and on the energy of the incoming photon. %Typical values for the 68\% containment radius are around 10$^\circ$ at 10 MeV and 1$-$5$^\circ$  at 100 MeV.

\subsection{Some examples of MeV $\gamma$-ray telescopes }

The Imaging Compton Telescope COMPTEL~\citep{COMPTELtelescope} was one of four instruments on-board NASA's Compton Gamma-Ray Observatory (CGRO)\footnote{\url{https://heasarc.gsfc.nasa.gov/docs/cgro/cgro.html}},  devoted to the exploration of the gamma-ray sky. CGRO operated from 1991 to 2000 orbiting Earth at an altitude of $\sim$450 km with a 28.46$^\circ$ orbital inclination. 
The COMPTEL telescope detected gamma rays with energies between 0.8 and 30 MeV via Compton interactions of the photons with two layers of scintillators. The energy resolution was 10\%$-$5\% (at 1 and 20 MeV, respectively). COMPTEL had a field of view (FoV) of $\sim$1.5 sr and covered the entire sky in 340 pointings with roughly 2 weeks exposure each. The exposure was not uniform across the sky as shown in Fig. 2 of Ref.~\cite{COMPTELexposure}.

Also on board CGRO, the Energetic Gamma Ray Experiment Telescope (EGRET)~\citep{EGRETtelescope} employed spark chambers to detected gamma rays between 20 MeV and  30 GeV that converted into a $e^+e^-$ pair within the instrument. The energy resolution was about 20\% over the central part of the energy range; EGRET effective area was about 1500 cm$^2$ between 200 MeV and 1 GeV, falling rapidly at higher and lower energies and with an approximately off-axis Gaussian profile with FWHM of $\sim$40 degrees. EGRET has been succeeded by the \textit{Fermi}-LAT above 100 MeV \citep{theLAT}.

The Compton Spectrometer and Imager (COSI)~\citep{COSIpos, COSIpaper} is a Compton telescope which will survey the sky between 0.2 and 5 MeV with a wide FoV ($\geq$25\% of the sky). The instrument is designed and optimized to detect spectral lines such as the 511 keV line from positron annihilation, and gamma-ray spectral lines from radioactive decays of isotopes produced by stellar nucleosynthesis, crucial probes of the evolution of massive stars and their supernova explosions. COSI will also measure polarization in the same energy range which will provide unique new insights into the underlying physics of astrophysical jets and other particle acceleration regions. In October 2021, COSI was selected by NASA to continue development as a Small Explorer mission, to be launched within the next 5 years. COSI will be key to accomplish the great science enabled by the observations in the MeV energy band\footnote{For more about the science enabled by MeV astrophysical observations we suggest the many white papers on the topic \citep[e.g.,][]{DeAngelis, Zoglauer2005thesis}.}. Before being developed as a space-based mission, COSI's technology has been tested in several balloon campaigns (see Tab.~\ref{tab:1}), which resulted in more than one month of data taking. COSI's selection as a future space-based explorer mission is of interest for the wide community of nuclear physicists and medium-gamma-ray astrophysicists.

Other proposed space-based missions to operate in the MeV gap include eASTROGAM~\citep{eASTROGAM}, and AMEGO~\citep{AMEGO}, which are examples of survey Compton-pair telescopes capable of measuring nuclear lines, continuum emission and polarization;  AMEGO-X~\citep{AMEGOXpos} is another example of all-sky survey Compton-pair telescope with a smaller volume than the previous two, optimized for continuum measurements and focused on the observation of multimessenger sources, such as gamma-ray bursts for joint detection with of gravitational waves and Galactic and extragalactic sources of high-energy cosmic rays and high-energy neutrinos\footnote{AMEGO-X will be proposed in the next NASA call for medium-size missions}; GECCO~\citep{GECCOpos} is a pointing Compton telescope optimized to observe the Galactic center and Galactic plane with the possibility to apply a coded aperture mask to greatly enhance its angular resolution and multiple-source discrimination power.

In Tab.~\ref{tab:1} we summarize some of the key characteristics of the missions and mission concepts mentioned above.\\

%\end{paracol}

\begin{table}[t]
\centering
    \small 
    \begin{tabular}{l|c|c|c|c|c|c}
Instrument & Detector type & Energy [MeV] &  FoV [sr] & Obs. Mode & Mission Size & Mission status \\
\hline
COMPTEL & Compton & 0.8 $-$ 30 & 1.5 & pointed & Great Obs. (CGRO) & Ended\\
EGRET & Pair & 20 $-$ 30000  & 0.4 & pointed & Great Obs. (CGRO) & Ended\\
\textit{Fermi}-LAT & Pair & 20 $-$ $>$3$\times10^5$  & $>$2 & survey & Space probe$^{(*)}$  & Surveying\\
COSI-balloon & Compton & 0.2 $-$ 5 & 3.1 & survey &  Balloon-borne &  Campaigns$^{(**)}$ \\
COSI & Compton & 0.2 $-$ 5 & 3.1 & survey & Space small & Launch in 2025 \\
eASTROGAM & Compton-Pair & 0.3 $-$ 3 & $>$2.5 & survey & Space medium & Concept \\
AMEGO & Compton-Pair & 0.2 $-$ 10000 & $>$2.5 & survey & Space probe & Concept \\
AMEGO-X & Compton-Pair & 0.3 $-$ 1000 & $>$2.5 & survey& Space medium & Concept \\
GECCO & Compton & 0.05/0.1 $-$ $\sim$10  & $\sim$1 & pointed & Space medium & Concept \\
\hline
    \end{tabular}
    \caption{Summary of some characteristics of previous, current, future and proposed imaging Compton and Compton-pair telescopes in the MeV band.\\
            $^{(*)}$ This classification is based on the modern nomenclature for the cost metrics of NASA's space missions, which has been defined after \textit{Fermi} selection.\\
            $^{(**)}$ COSI 2016 (Wanaka, New Zealand), COSI 2014 (McMurdo Station, Antarctica).}

    \label{tab:1}
\end{table}
%\begin{paracol}{2}
%\linenumbers
%\switchcolumn

%%%%%%%%%%%%%%%%%%%%%%%%%%%%%%%%%%%%%%%%%%
\section{Source sensitivity of MeV telescopes}
\label{sec:sensitivity}
\subsection{Background estimation}
\label{subsec:bkg}
\begin{figure}
    \centering
    \includegraphics[scale=0.46]{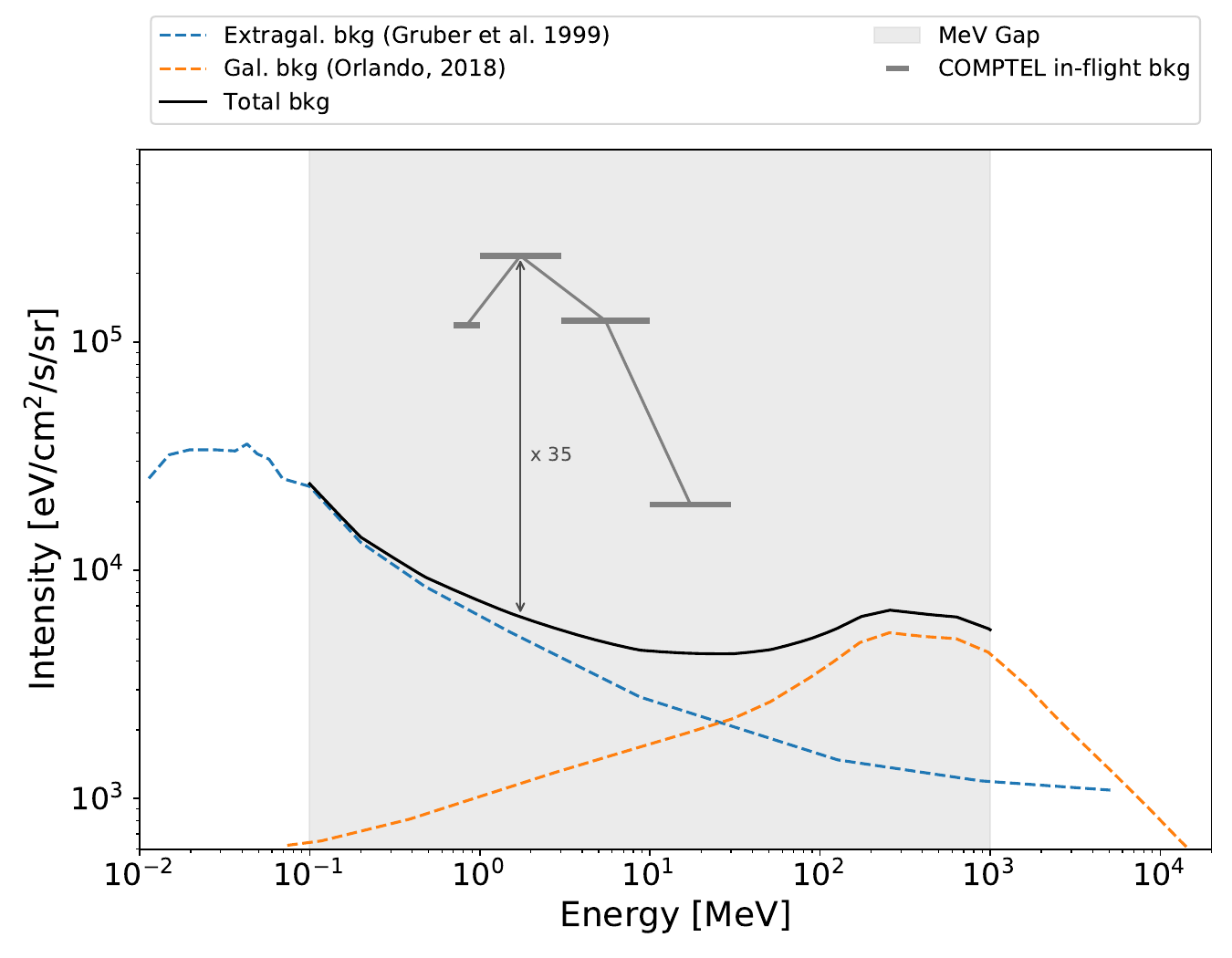}
    \caption{Galactic (dashed light blue line), Extragalactic (dashed orange line) astrophysical background components and the sum of the two (solid black line). The gray solid line illustrates the in-flight measure background by COMPTEL as reported in Ref.~\cite{COMPTELtelescope} scaled for a region of 1 sr extension (see text for details). }
    \label{fig:bkg-Aeff}
\end{figure}

For prediction of the detection sensitivity of an instrument to a given source, a crucial element is a realistic estimate or a measurement of the background. 
As an example, we show the background flux measured by COMPTEL in Fig.~\ref{fig:bkg-Aeff}, in comparison to the mid-latitude-Galactic and extragalactic background components, for which we provide more details in the next paragraph. The COMPTEL background flux was derived from the in-flight background counts reported in Table 12 of Ref.~\cite{COMPTELtelescope}: note that these background counts correspond to the specific event selection declared in section 8.1 of \cite{COMPTELtelescope}. We derive the flux by dividing the counts by the effective exposure reported in Table 13 of that reference for the same energy bins, and scaled for the radius of a 1 steradian region (see Appendix~\ref{app:b}). The measured COMPTEL background is up to a factor of 35 greater than the total astrophysical gamma-ray intensity: it is evident how the instrumental background is an important consideration for Compton telescopes, and in the case COMPTEL it is the dominant source of background.

A guaranteed astrophysical background from the FB region is represented by the diffuse astrophysical background, which has two components: the extragalactic gamma-ray background and the Galactic diffuse emission. We are assuming that the contribution from point sources within the FB region is known, sufficiently well modelled and subtracted \citep[and/or masked away following the technique used, in the pair-production regime, by][]{FBLAT}. Once the instruments are in orbit, we expect that they will detect and characterize the point-like sources, which will allow this approach to be used. We note, however, that any mis-modeling of the detected sources would contaminate the FB measurement. The extragalactic background can be assumed isotropic and we consider, for our simulations, the measured spectrum of diffuse cosmic hard X-rays by High Energy Astronomical Observatory 1 \citep{Gruber1999}. The Galactic diffuse emission is not isotropic, depending strongly on Galactic latitude, especially below $|b|<10^\circ$. Here, we consider the spectrum of the total Galactic diffuse emission at intermediate latitudes ($10^\circ<|b|<20^\circ$) estimated by \cite{Orlando} and for simplicity we consider it to be isotropic within the FB region.  This assumption is a conservative choice to estimate the sensitivity to detect the FB: the measured total diffuse emission is less bright at higher latitudes which would result in better sensitivity. However, an accurate model of the Galactic diffuse emission morphology is expected to be achievable for the actual measurement of the MeV FB emission with flight data, with contributions from components whose intensities are currently uncertain, such as gamma-ray emission associated with the Loop~I and Loop~IV radio filaments, being well determined by the data. With these considerations, systematic uncertainties of the model for the non-FB Galactic interstellar emission will not be a limiting factor for the sensitivity of the measurement.

Fig.~\ref{fig:bkg-Aeff} (left) shows the astrophysical diffuse emission discussed above. On top of the astrophysical background, MeV gamma-ray detectors typically suffer large instrumental backgrounds which dominate the trigger rates on time scales longer than a few minutes. Such backgrounds include contributions from gamma rays emitted by the Earth's atmosphere, charged particles (cosmic rays) passing through the detector and radioactive decays from activation induced by the latter. The charged-particle background can be reduced by surrounding the instrument with plastic scintillator plates or similar sensors that can detect and veto the passage of charged particles. The activation background, which is mostly irreducible, is composed of radioactive decays within the instrument due to activation of the detector material caused by cosmic-ray interactions (dominant in the UC and TC regimes), and secondary gamma rays produced by the impact of cosmic rays in the outer layers of the spacecraft (e.g. the micrometeoroid shielding), before the vetoing scintillators. The irreducible instrumental backgrounds strongly depend on the spacecraft design, materials and orbital altitude and inclination, and need to be carefully evaluated through detailed simulations and in-flight measurements.

\subsection{Prescription to derive MeV telescope sensitivity to the FB}
\label{subsec:prescription}
We define the FB region as a circular region with a radius $R_{FB}=20^\circ$ covering the intermediate latitudes with centers at $l=0$ and b$=\pm 30^\circ$. The 3$\sigma$ sensitivity in a given energy bin is obtained by estimating the total number of counts needed to have a signal-to-noise ratio SNR=3 given the total number of background events expected while observing the FB region:
\begin{equation}
    {\rm SNR} = \frac{N^{\Delta E}_{{\rm sig}}}{\sqrt{N^{\Delta E}_{{\rm sig}} + N^{\Delta E}_{{\rm bkg}}}}
    \label{eq:eq1}
\end{equation}

% In the UC regime, the number of background counts in a given energy range can be obtained through simulations (for a given instrument design) within an extraction radius, namely the half width at half maximum of the ARM distribution, of the size in the FB region. 
In the P event regime, besides running simulations, it is possible to estimate the background counts from the expected background flux in photons per unit area, time, solid angle in the energy bin of interest, $I^{\Delta E}_{\mathrm{bkg}}$, and correcting by the effective area and the observation time of the telescope:
\begin{equation}
    \label{eq:bkg}
    N^{\Delta E}_{{\rm bkg}} = t_{{\rm obs}} ~ I^{\Delta E}_{\mathrm{bkg}}~ A^{\bar{\Delta E}}_{\mathrm{eff}, \bar{\Theta}} ~ \Omega
    %\int_0^{20^\circ} \mathrm{sin}(\theta) d\theta
\end{equation}
where $A^{\bar{\Delta E}}_{\mathrm{eff}}(\theta)$ is the average effective area of the telescope in the energy range considered and off-axis angle range $\Theta$ within the field of view (FoV); with $\Omega$ being the the solid angle of the FB region ($\Omega=2\pi(1-\cos(R))\sim 0.38$ sr).
The observation time for wide-FoV survey-like telescopes can be estimated as $t_{{\rm obs}}=t_{{\rm mission}}\frac{{\rm FoV}}{4\pi}f$, where $f$ is a factor that takes into account possible dead times due to Earth occultation or temporal survey suspensions (e.g., due to targeted pointing or south Atlantic anomaly crossing). For a pointing instrument one can assume to point to the center of one of the FB at a time and take into account the dependence of the effective area on the off-axis angle (the angle between the boresight of the instrument and a position in the sky) within the FB region with the following substitution 

\begin{equation}
A^{\bar{\Delta E}}_{\mathrm{eff}, \bar{\Theta}} ~ \Omega ~~\rightarrow~~ 2\pi\int_0^{20^\circ} A^{\Delta E}_{\mathrm{eff}}(\Theta) \mathrm{sin}(\Theta) d\Theta 
\end{equation}

The number of signal counts, $N_{{\rm sig}}$ needed to achieve a 3$\sigma$ detection given N$_{bkg}$ can be obtained by requiring ${\rm SNR}=3$ in Equation \ref{eq:eq1}, and the flux sensitivity for P events can be obtained as:

\begin{equation}
    F^{\Delta E}_{{\rm SNR}=3} = \frac{N^{\Delta E}_{{\rm sig}}}{t_{{\rm obs}} ~ A^{\bar{\Delta E}}_{\mathrm{eff}, \bar{\Theta}}}~[{\rm cm}^{-2}{\rm s}^{-1}]
    \label{eq:sig}
\end{equation}

For UC events, if the point source sensitivity is known, it is possible to determine the sensitivity for the FB region by means of the scaling of the sensitivity with the square root of the background event count (for the case of a background-dominated source), or equivalently with the power of 1/4 of the extended source area (see Appendix \ref{app:b}):

\begin{equation}
    F^{\Delta E}_{{\rm SNR}=3} = F^{{\rm PS},\Delta E}_{{\rm SNR}=3}  \left(\frac{\Omega}{\pi R_{{\rm ARM}/2}^2}\right)^{1/4}
    \label{eq:sigUC}
\end{equation}

where $R_{ARM/2}$ is HWHM of the ARM distribution in steradians for the energy bin considered. Note that the angular extension $\Omega$ in Eq.~\ref{eq:sigUC} is limited by the angular resolution of the instrument.

For TC events, the scaling in sensitivity with the extension of the source should be somewhere in between the softer scaling for UT events and the stronger scaling for P events. To estimate such scaling is beyond the purpose of this paper. Hence in this work, unless stated otherwise, we treat TC events as P events adopting Eq~\ref{eq:bkg} to estimate the sensitivities; namely we are assuming the worst-case scenario in which TC sensitivities scale like P sensitivities with the extension of the source\footnote{Eq~\ref{eq:bkg} still holds for TC events in the case it is possible to select events (which look like arcs in the sky) coming from the region of interest without losing too much effective area. This selection can be done, for example, by discarding all events whose arc is not fully contained in the region of interest, or, more loosely, keeping all events whose arc falls within the region for more than, say, 95\% of its extension. Intuitively, this is the case in which the extension of the arc for the majority of the events does not exceed the size of the FB region.}.

The sensitivity to the pair of bubbles, assuming that they have identical emissions, is obtained by combining the sensitivity to one bubble as $I_{2FB} = \frac{1}{\sqrt{2}}I_{1FB}$.

\subsection{COMPTEL, COSI, and AMEGO-X mission concept sensitivities}
\begin{figure}
    \centering
    \includegraphics[width=12cm, height=10cm]{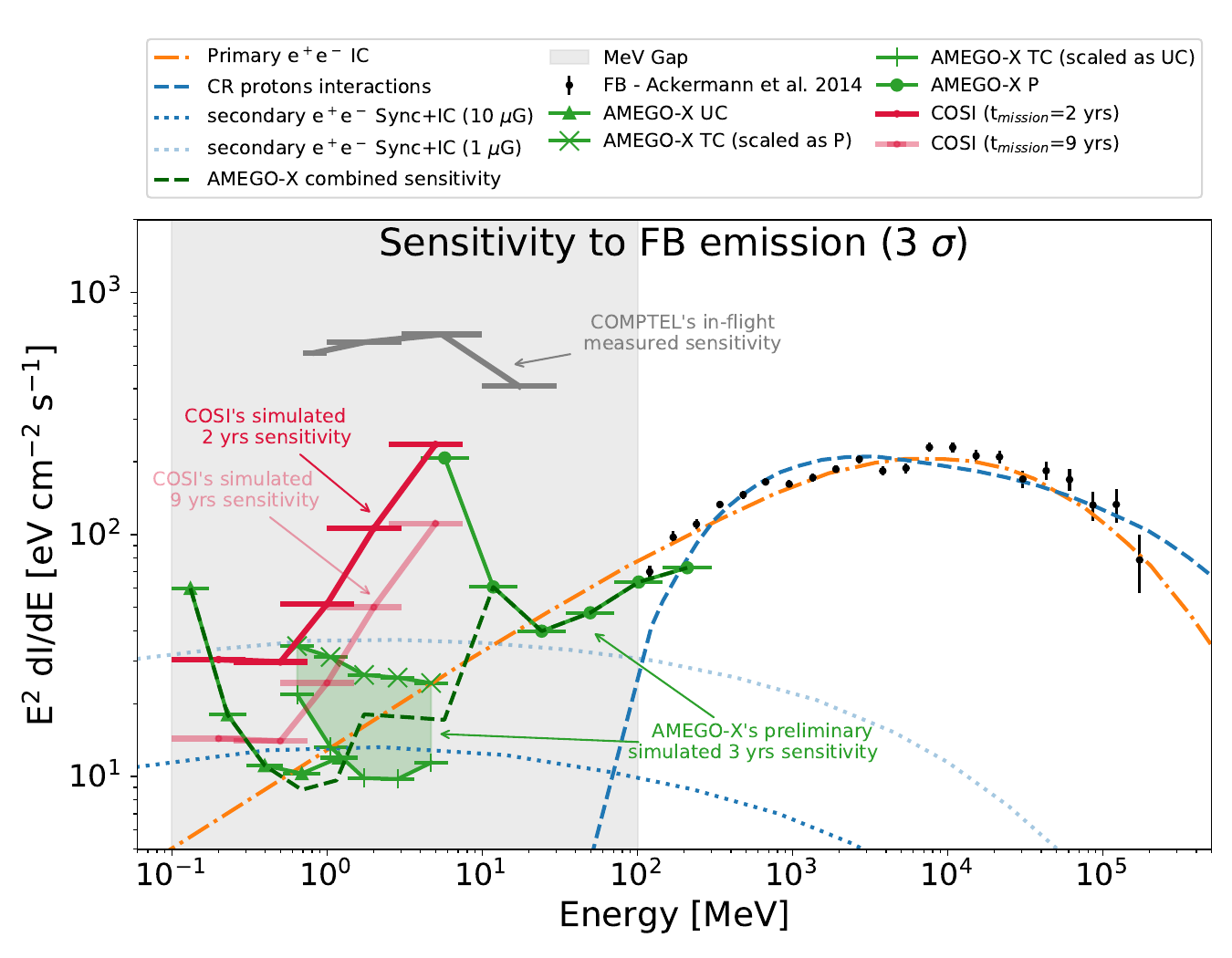}
    \caption{Combined FB sensitivity for both bubbles in the MeV gap for COSI, COMPTEL and AMEGO-X. COMPTEL sensitivity is derived from the Table 13 of Ref.~\cite{COMPTELtelescope} and scaled for the FB extension as detailed in the text. COSI sensitivity has been provided by the COSI team and are based on Geant4 \citep{Geant4} simulations using the COSI design requirement performance for a disk-like source with 20-degree radius. AMEGO-X sensitivity for UC events has been obtained scaling the point source sensitivity as detailed in the text; the two sensitivity curves in the TC regime have been derived by treating TC events as UC ones (green ``+'') or as P events (green ``x''). Both are based on preliminary background simulations run with an ARM cut of 20$^\circ$ simulations. The combined AMEGO-X sensitivity is reported (dark-green dashed line), where for UC events we considered the median of the sensitivity curves obtained by treating TC events as UC and as P events. The orange line show the IC emission that best fit the LAT measured FB spectrum in \cite{FBLAT}. The blue lines show primary and secondary $\gamma$-ray emission from hadronic processes. The shaded gray region marks the MeV gap. }
    \label{fig:MeVzoom}
\end{figure}

COMPTEL was designed to operate in the UC regime, so we followed the procedure for the UC regime described in the previous section to scale the point source continuum sensitivity of COMPTEL to a 1 steradian extended region.  We use the point source continuum sensitivity reported in Table 13 of Ref.~\cite{COMPTELtelescope} and the 1-$\sigma$ ARM width illustrated in Figure 28 of the same paper. Such a sensitivity is for a 2 weeks exposure, hence, knowing that the sensitivity scales as the squared root of the time, we obtain the sensitivity for the average exposure within the FB region as reported in Ref.~\cite{COMPTELexposure}: the average exposure in the FB region for the whole mission is about 55 days. In Fig.~\ref{fig:MeVzoom} we show that the sensitivity of COMPTEL is simply insufficient to detect the FB. 

COSI's sensitivity curves are illustrated in Fig.~\ref{fig:MeVzoom}\footnote{COSI sensitivity curves have been made available through a private communication with John Tomsick, PI of the COSI mission.}. We report the 3-$\sigma$ sensitivity curves for the nominal 2 years of the mission (red curve) and for 9 years (light red curve). During its nominal lifetime of 2 years, COSI would observe the FB region for approximately 6 months and detect an hadronic component from the FB (for magnetic field values $\lesssim 1\mu$G). Interesting upper limits could be set below $\sim$1 MeV extending the mission beyond the nominal lifetime.

We also report the case of the AMEGO-X Compton-pair telescope concept in Fig.~\ref{fig:MeVzoom}. The UC, TC and P event regimes are shown in green, and have been obtained following the procedure described in Sec.~\ref{sec:sensitivity}. In particular, the sensitivity for the UC regime has been derived using Eq.~\ref{eq:sigUC} from the point-source continuum sensitivity estimated through simulations\footnote{AMEGO-X point-source sensitivities have been carried out by the AMEGO-X team with MEGALib \citep{MEGALib}. For TC and P regimes we used Eq.~\ref{eq:sig} to derive the sensitivities, where preliminary background simulations and simulated instrument effective area have been provided by the AMEGO-X team\footnote{\url{https://asd.gsfc.nasa.gov/amego-x/}}.}
The sensitivities are based on simulations of the expected background rates and gamma-ray effective area of AMEGO-X, using a preliminary detector model. The background simulations, similarly to the ones use for COSI, include contributions from cosmic rays (protons, alpha particles, electrons), instrument activation, atmospheric background, and the extragalactic (isotropic) gamma-ray background. More refined simulations will only become available when the instrument design is finalized. Would the mission proceed, the background rates will be derived from actual in-flight measurements for which an accuracy well under 10\% is expected.
AMEGO-X is designed to operate in survey mode, with a wide FoV of 2.5 sr in the P regime and up to 3.6 sr in the Compton regime. Fig.~\ref{fig:MeVzoom} shows that an instrument like AMEGO-X would be able to detect a pure-leptonic emission from the FB almost in the entire energy range of the MeV gap in a 3-year sky survey\footnote{Three years is the nominal duration for NASA's Medium Explorer missions.}, and could detect hadronic (secondary leptonic) emission in an energy range that depends on the magnetic field present within the FB. In the case of hybrid scenarios, the detection of a spectral break in the MeV energy range would constrain the relative contribution of the leptonic component over the hadronic one populating the FB, and, once the primary hadronic population is fixed, the intensity of the secondary leptonic emission would constrain the magnetic field within the bubbles.\\

The sensitivity estimations presented in this section come with some caveats. First, the uniformity of the FB is an approximation, and some level of non-uniformity in the intensity distribution and spectrum might affect the sensitivity estimates (e.g., and increase the signal-to-noise ratio in the brighter regions). Second, as already mentioned, the estimation of the background components is based on simulations (except for the case of COMPTEL). Once an instrument is in orbit, the instrumental background can be calibrated, and measurements of the actual astrophysical background and sources can be obtained within a certain level of accuracy (expected to be of the order of a few \% or lower, based on similar, previously developed instruments). 
For example, the internal activation background can be modeled based on the strength of detected nuclear lines, shields around the instrument reject the charge particle background, and the upward moving atmospheric background photons can be characterized by occasionally pointing the instrument at the atmosphere. In addition, COSI and AMEGO-X will have several additional particle and background detectors on board to monitor the space radiation environment and enable their accurate characterization. As a consequence, we expect to be able to characterize the background with an uncertainty significantly below 10\% (the exact number depends on many factors and will only be known after launch).
Finally, our computation of the AMEGO-X sensitivity, does not include any event quality cut which would reduce the background contamination and, hence, increase the signal-to-noise ratio. In Appendix \ref{app:c}, as an example we show how the AMEGO-X FB sensitivity changes when scaling the background rate, and we present a simple procedure for estimating the variation of the sensitivity admitting a certain level of uncertainty on the simulations.

%\end{paracol}

\section{Conclusions}
\label{sec:concl}
In this work we showed the potential for observations of MeV $\gamma$ rays to solve the long-standing mystery behind the origin of the FB emission, which is a crucial aspect to understand how these apparently giant structures themselves originated. 

Firstly we noted how pure-leptonic and pure-hadronic emission scenarios diverge in the MeV gap.  Hence we discussed the main characteristics of some instruments operating in this energy range. In particular we computed the sensitivity of some past and future MeV telescopes. We showed that COMPTEL was not sensitive enough to detect the predicted emission from the FB, while COSI, recently selected by NASA's Small Explorer missions program, could detect a hadronic component at 0.1$-$0.6 MeV already within its prime mission time (2 years), or could set upper limits on the FB emission in this energy range, constraining the magnetic field strength within the bubbles for hadronic scenarios. Additionally, we illustrate the concept of AMEGO-X as an example of a future wide-FoV survey Compton-pair telescope: according to preliminary simulations, AMEGO-X would be sensitive to a pure-leptonic component from the FB in almost the entire energy range of the MeV gap (100 keV $-$ 100 MeV), after three years of survey. 

Additionally we noted how, in the case of a hybrid scenario, in which both a leptonic and a hadronic component to the emission of the FB, the observation of a spectral break in the MeV band would constrain the relative contribution of the two populations. Additionally the intensity of the secondary leptons generated in hadronic interactions would constrain the magnetic field within the FB. This scenario would be accessible to a wide-FoV survey Compton-pair telescope sensitive to the FB emission throughout almost the entire MeV gap energy range.

In conclusion, MeV $\gamma$-ray observations will be able to resolve the emission mechanisms and origin of the Fermi Bubbles. The future Compton telescope COSI, on top of the primary science it has been designed for, will shed some light on the main emission mechanism from the FB. A Compton-pair survey telescope such as AMEGO-X have the potential to ultimately unveil the origin of the FB emission.

%% IMPORTANT! The old "\acknowledgment" command has be depreciated. It was
%% not robust enough to handle our new dual anonymous review requirements and
%% thus been replaced with the acknowledgment environment. If you try to 
%% compile with \acknowledgment you will get an error print to the screen
%% and in the compiled pdf.
\begin{acknowledgments}
M.N. acknowledges John Tomsick and the COSI team for providing the COSI's continuum sensitivities, and Eric Burns for the fruitful discussions and for encouraging writing this manuscript. We acknowledge Regina Caputo and the AMEGO-X team for the work done on the simulations and instrument performance. The material is based upon work supported by NASA under award number 80GSFC21M0002. 
\end{acknowledgments}

%% To help institutions obtain information on the effectiveness of their 
%% telescopes the AAS Journals has created a group of keywords for telescope 
%% facilities.
%
%% Following the acknowledgments section, use the following syntax and the
%% \facility{} or \facilities{} macros to list the keywords of facilities used 
%% in the research for the paper.  Each keyword is check against the master 
%% list during copy editing.  Individual instruments can be provided in 
%% parentheses, after the keyword, but they are not verified.

% \vspace{5mm}
% \facilities{HST(STIS), Swift(XRT and UVOT), AAVSO, CTIO:1.3m,
% CTIO:1.5m,CXO}

%% Similar to \facility{}, there is the optional \software command to allow 
%% authors a place to specify which programs were used during the creation of 
%% the manuscript. Authors should list each code and include either a
%% citation or url to the code inside ()s when available.

% \software{astropy \citep{2013A&A...558A..33A,2018AJ....156..123A},  
%           Cloudy \citep{2013RMxAA..49..137F}, 
%           Source Extractor \citep{1996A&AS..117..393B}
%           }

%% Appendix material should be preceded with a single \appendix command.
%% There should be a \section command for each appendix. Mark appendix
%% subsections with the same markup you use in the main body of the paper.

%% Each Appendix (indicated with \section) will be lettered A, B, C, etc.
%% The equation counter will reset when it encounters the \appendix
%% command and will number appendix equations (A1), (A2), etc. The
%% Figure and Table counter will not reset.

\appendix

\section{Pure Leptonic and pure Hadronic models}
\label{app:a}
A careful morphological and spectral characterization of the FB was performed by the Fermi-LAT collaboration in Ref.~\cite{FBLAT}. In that work,  pure leptonic and a pure hadronic scenarios evaluated to describe the average FB spectrum. We consider the best-fit models for the two scenarios proposed by Ref.~\cite{FBLAT}, and look at the expected gamma-ray emission in the MeV band. 

In Ref.~\cite{FBLAT}, the spectrum of the cosmic-ray electron (CRe) population for the pure leptonic model  was assumed to be a power law with an exponential cut off $\propto E^{-n}e^{E/E_{{\rm cut}}}$, where the best-fit parameters where found to be $n=2.17\pm0.05$ and $E_{{\rm cut}}=1.25\pm0.13$ TeV. This population of electrons and positrons interacts with the ISRF which was modelled as in the GALPROP code \citep{strong2009galprop}. The resulting IC spectrum is reported in Fig.~\ref{fig:landscape} as a dot-dashed orange line. The less-energetic side of the CRe population also interacts with the magnetic field present inside the bubbles and produces synchrotron emission; a magnetic field intensity of $B=8.4\pm0.2~\mu$G\footnote{It is worth mentioning the systematic error associated with the best-fit magnetic field intensity as quoted by \cite{FBLAT}: $B=8.4\pm0.2[{\rm stat}]^{+11.2}_{-3.5}[{\rm syst}]~\mu$G. } has been found to reproduce the microwave haze quite well (see dashed orange line in Fig.~\ref{fig:landscape}).

As for the hadronic model adopted in Ref.~\cite{FBLAT}, the assumed cosmic-ray proton (CRp) spectrum was a power law with exponential cut off $\frac{dn(p)}{dp}\propto p^{-n}e^{pc/E_{{\rm cut}}}$, with best-fit values being $n=2.13\pm0.01$ and $E_{{\rm cut}}=14\pm7$ TeV, and assuming a ionized hydrogen column density of $n_H=0.01 cm^{-3}$. The primary hadronic gamma-ray emission due to $\pi^0$ decay is shown in Fig.~\ref{fig:landscape} as a blue wide-dashed line, while the gamma-ray emission from the population of secondary leptons produced through hadronic interactions is reported as dotted lines for different values of assumed magnetic field ($1\mu$G and $10\mu$G in light and darker blue respectively). It is evident that the synchrotron emission from the secondary leptons is not enough to reproduce the microwave haze (further comments about this can be found in Sec.7.2 of Ref.~\cite{FBLAT}).

\section{Untracked Compton events scaling with source extension}
\label{app:b}

\begin{figure}
    \centering
    \includegraphics[angle=-90, width=12cm]{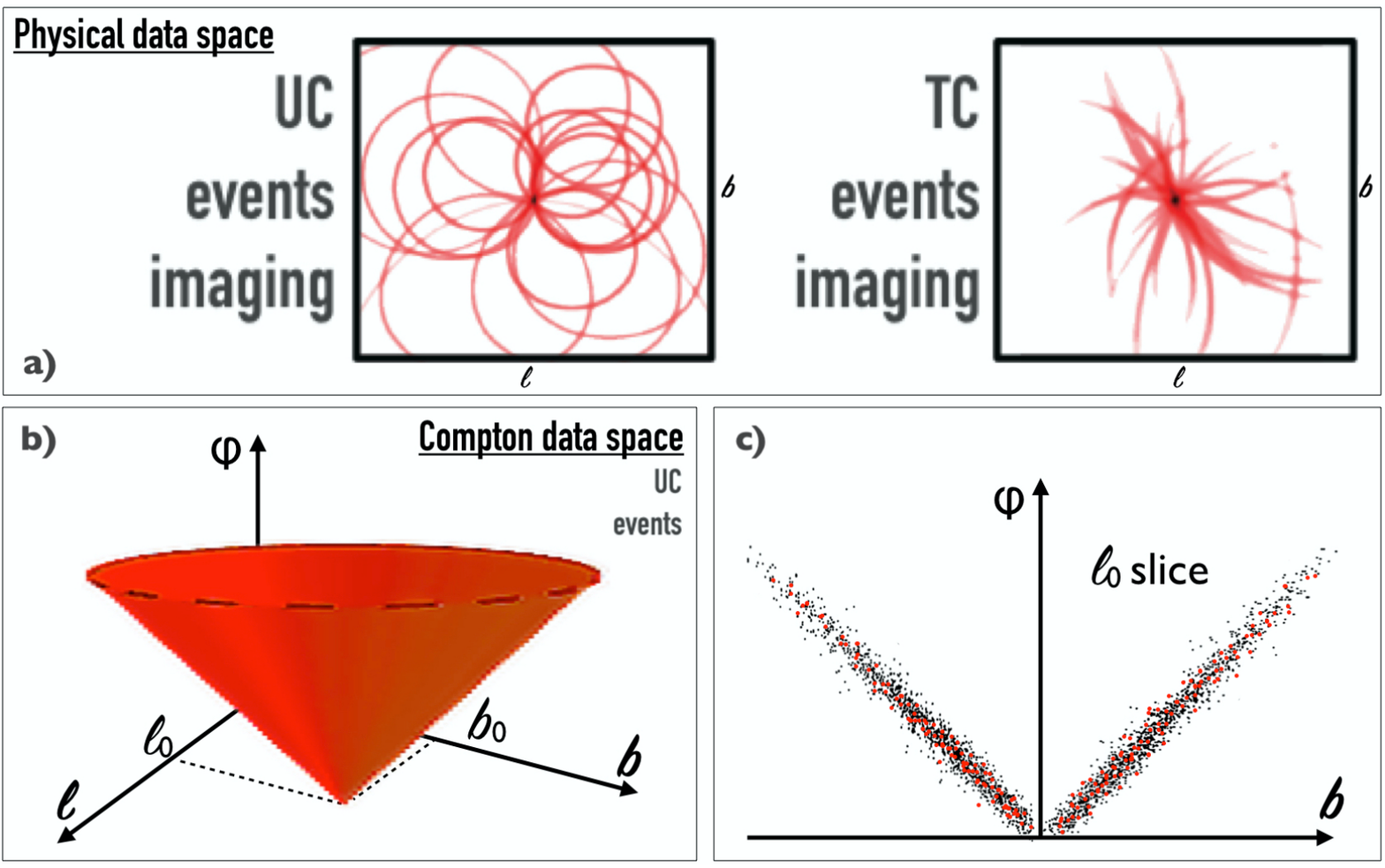}
    \caption{a) Example of the imaging of a point source in Galactic coordinates in the untracked Compton (UC) regime and in the tracked Compton (TC) regime. b) Events cone in Compton data space ($\phi$, $\ell$, $b$). c) Median section of the Compton data space cone: the dispersion of the data points from the geometrical cone defines the width of the ARM distribution. The effect of the extension of a source on the Compton cone is to ``thicken'' the cone mantle by approximately the extension radius.}
    \label{fig:compton}
\end{figure}

In Sec.~\ref{sec:MeVtelescopes} we briefly described the typical procedure for imaging with Compton events by back projecting onto the sky the Compton circles (UC regime) or event arcs (TC regime). A schematic illustration of the imaging of a point source for both UC and TC regimes is given in Fig.~\ref{fig:compton} (a). 
It is clear, for UC events in particular, how the dominant background events can contaminate the image and wash out the signal from the source, and a selection based exclusively on the event circle locations in the 2D sky coordinates space could result in a significant reduction of the effective area and worse sensitivities. So, when it comes to event selection and data analysis, a more convenient space to work with is the Compton data space (CDS), a 3D space whose coordinates are the direction of the axis of the event Compton circle (in sky coordinates) and the Compton scattering angle: each UC event is a point in this space. In the case of a perfect instrument (infinite energy resolution) and full absorption of the Compton-scattered photon in the second layer of the detector, the geometrical place defined by the events coming from a source in the sky ($\ell_0$, $b$) is a cone with a 90$^\circ$ opening angle (Fig.~\ref{fig:compton} (b)). If the Compton-scattered photons are not fully absorbed and the instrument has a limited energy resolution, the events in the CDS are not perfectly on the cone but spread out from the surface of the cone, ``thickening'' the cone mantle: the distribution of the angular distance of the event points from the ideal cone is the ARM, whose FWHM gives the angular resolution of the instrument. Events from an extended source would result in a further thickening of the cone mantle to the size of the source radius extension. Assuming a disk-like extended emission and working in the Cartesian approximation of the cone, the number of background events that would be counted by selecting the events that fall in the cone mantle within a given ARM value is proportional to the volume:

\begin{equation}
    N_{{\rm bkg}} \propto V_{{\rm Cone~mantle}} = \frac{\pi}{3}h(h+R)^2 - \frac{\pi}{3}h(h-R)^2 = \frac{2}{3}\pi h^2R
\end{equation}

where $h$ is the height (equal to the radius) of the cone in the CDS, and $R$ is the half-thickness of the mantle (aka the extraction radius of the source). Recalling that the flux sensitivity ($I_\sigma$) scales with the square root of the background counts (in the background dominated regime) we have:

\begin{equation}
    \frac{I_\sigma^{ES}}{I_\sigma^{PS}} \propto \frac{\sqrt{V_{{\rm Cone~mantle}}^{ES}}}{\sqrt{V_{{\rm Cone~mantle}}^{PS}}} = \sqrt{\frac{R^{ES}}{R^{PS}}} = \left(\frac{S^{ES}}{S^{PS}}\right)^{\frac{1}{4}}
\end{equation}
 where the labels $PS$ and $ES$ denote a point source and extended source respectively; $R^{PS}$ is the HWHM of the ARM distribution while $R^{ES}$ is the radius of the disk that approximates the source extended emission; $S^{ES}$ is the area of the extended source (disk approximation), while $S^{PS}\sim\pi {\rm ARM}_{HWHM}^2$.

\section{Sensitivity scaling with the background}
\label{app:c}
\begin{figure}
    \centering
    \includegraphics[width=12cm]{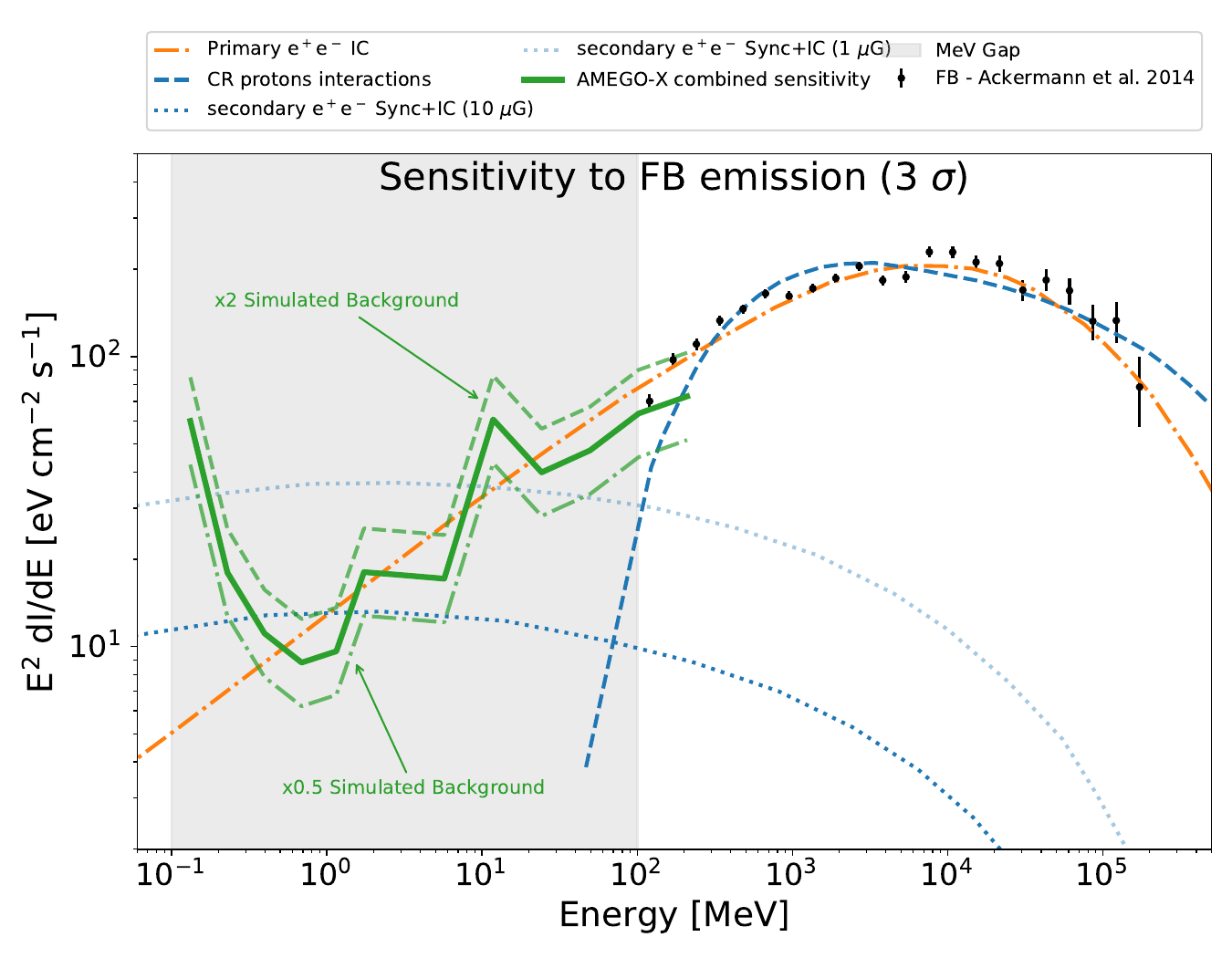}
    \caption{Variation of the AMEGO-X combined (3 years) sensitivity if assuming a factor of 2 more and less background rate. the solid green line correspond to the dashed dark green line in Fig.~\ref{fig:MeVzoom}.}
    \label{fig:BKGscaling}
\end{figure}

In a background-dominated regime, as the case considered in this study, the source sensitivity, given a fixed SNR, scales proportionally to the square root of the background counts. Using this relation, in Fig.~\ref{fig:BKGscaling}, we compare the AMEGO-X combined (UC+TC+P events) sensitivity with that obtained with a simulated background scaled by factors of 2 and 0.5. In the extreme case where the simulated background is enhanced by 100\% the AMEGO-X sensitivity would lie just above the predicted intensity of the leptonic model. Therefore mis-modeling the background intensity by anything less than 50\% would not affect our conclusions.

Below 30 MeV, the background is dominated by the extragalactic astrophysical component plus the irreducible (what remains after well defined quality cuts) instrumental component (cosmic rays, activation, atmospheric gamma-ray photons). Once on-orbit, we will be able measure the rate and energy spectrum of the background very well using ``off regions'' with no gamma-ray source, as it is isotropic after earth horizon cuts. To estimate how well we might be able to measure this background, we can look at other, similar gamma-ray  instruments. Let's consider the Fermi LAT isotropic background template. The flux uncertainty in the lowest bin provided (starting at 30 MeV) improved from 7\% (pass 6 R3, after one year on orbit) to 0.1\% (pass 8 R3, after 8 years on-orbit)\footnote{ numbers from \url{https://fermi.gsfc.nasa.gov/ssc/data/access/lat/BackgroundModels.html}}. Therefore, we believe that we will be able to measure the isotropic background for AMEGO-X with similar accuracy.

As mentioned in the main text, by applying some event-quality cuts, we will be able to enhance the signal-to-noise ratio for the observation of the FB emission. Once such cuts will be defined and applied, we would be able to determine the variation of the AMEGO-X sensitivity when admitting a given level of uncertainty on the background component. This estimation can be done by simply redefining Eq.~\ref{eq:eq1} as follow:

\begin{equation}
{\rm SNR} = \frac{N^{\Delta E}_{{\rm tot}}-\alpha N^{\Delta E}_{{\rm bkg}}}{\sqrt{N^{\Delta E}_{{\rm tot}}}} = \frac{N^{\Delta E}_{{\rm sig}} + (1- \alpha)N^{\Delta E}_{{\rm bkg}}}{\sqrt{N^{\Delta E}_{{\rm sig}} + N^{\Delta E}_{{\rm bkg}}}}
\end{equation}

where $\alpha=1\pm{\rm syst.}$ (where syst. stands for the systematic uncertainty).

%% For this sample we use BibTeX plus aasjournals.bst to generate the
%% the bibliography. The sample631.bib file was populated from ADS. To
%% get the citations to show in the compiled file do the following:
%%
%% pdflatex sample631.tex
%% bibtext sample631
%% pdflatex sample631.tex
%% pdflatex sample631.tex
\clearpage

\bibliography{sample631}{}
\bibliographystyle{aasjournal}

%% This command is needed to show the entire author+affiliation list when
%% the collaboration and author truncation commands are used.  It has to
%% go at the end of the manuscript.
%\allauthors

%% Include this line if you are using the \added, \replaced, \deleted
%% commands to see a summary list of all changes at the end of the article.
%\listofchanges

\end{document}